\documentclass[twocolumn,10pt]{article}

\usepackage[T1]{fontenc}
\usepackage[utf8]{inputenc}
\usepackage{microtype}
\usepackage{colortbl}
\usepackage{xcolor}
\usepackage{graphicx}
\usepackage{amsmath,amssymb}
\usepackage{booktabs}
\usepackage[breaklinks]{hyperref}

\title{ACES: Accent Subspaces for Coupling, Explanations, and Stress-Testing in Automatic Speech Recognition}

\author{Swapnil Parekh\\ \normalsize Intuit}
\date{}

\begin{document}
\maketitle

\begin{abstract}
ASR systems exhibit persistent performance disparities across accents, but whether these gaps reflect superficial biases or deep structural vulnerabilities remains unclear. We introduce \textbf{ACES}, a three-stage audit that extracts accent-discriminative subspaces from ASR representations, constrains adversarial attacks to them, and tests whether removing them improves fairness. On Wav2Vec2-base with seven accents, imperceptible perturbations (${\sim}$60~dB SNR) along the accent subspace amplify the WER disparity gap by nearly 50\% (21.3$\to$31.8~pp), exceeding random-subspace controls; a permuted-label test confirms specificity to genuine accent structure. Partially removing the subspace worsens both WER and disparity, revealing that accent-discriminative and recognition-critical features are deeply entangled. ACES thus positions accent subspaces as powerful fairness-auditing tools, not simple erasure levers.
\end{abstract}


\section{Introduction}

{\sloppy
ASR systems have reached high accuracy on standard benchmarks yet exhibit substantial performance disparities across speaker groups~\cite{koenecke2020racial,wu2022accent,prasad2020accents}, which limit the accessibility of voice-driven technologies. Prior work primarily measures group WER gaps, probes where accent is decodable, or proposes training-based mitigation~\cite{ludecke2022residual}. We propose \textbf{ACES}, a representation-centric audit for accent disparity in ASR that uses an accent subspace as a \emph{test instrument} to constrain stress tests, measure aggregate vulnerability, and evaluate whether inference-time removal of that subspace changes disparity.

Our contributions are threefold. We introduce a three-stage audit (subspace extraction $\to$ subspace-constrained attacks with matched controls $\to$ project-out intervention) and present evidence on Wav2Vec2-base with seven accents that accent-subspace-aligned perturbations consistently amplify disparity beyond random-subspace controls---an effect confirmed to be specific to real accent structure via a permuted-label specificity control. We also report a negative intervention result: attenuating decodable accent can worsen disparity, which motivates caution in using linear ``erasure'' as a fairness fix. Section~\ref{sec:related} reviews related work; Section~\ref{sec:method} describes the method; Sections~\ref{sec:experiments}--\ref{sec:discussion} present experiments and discussion.
\par}

\section{Related Work}
\label{sec:related}

\subsection{Accent disparity and mitigation in ASR}

Accent mismatch is a major source of WER disparity~\cite{koenecke2020racial,wu2022accent}. Benchmarks such as AESRC~\cite{accents2020aesrc} and fairness-focused datasets (e.g., FaiST~\cite{jahan25faist}, ASR-FAIRBENCH~\cite{rai25asrfairbench}) provide multi-accent evaluation. Prasad and Jyothi~\cite{prasad2020accents} and Huang et al.~\cite{huang2023accent} probed where accent is decodable in ASR and accent identifiers. Mitigation often relies on training or adaptation (e.g., residual adapters~\cite{ludecke2022residual}). Unlike prior probing or training-based approaches, ACES audits an \emph{existing} model using accent subspaces as \emph{test instruments}: constrained perturbations and representation interventions to quantify and explain disparity.

\subsection{Adversarial robustness in ASR}

Adversarial examples can fool ASR~\cite{qin2019imperceptible,carlini2018audio}; Wav2Vec2~\cite{baevski2020wav2vec2} is vulnerable, and adversarial training can worsen demographic disparity~\cite{xu2021fairness}. Mechanism-focused attacks (e.g., subspace-directed triggers~\cite{singla2022minimal}) align with our goal: we constrain perturbations to the learned accent subspace at matched L2 budget to test whether perturbing that direction exacerbates disparity relative to a random-subspace control.

\subsection{Interpretability and representation intervention}

Concept erasure and null-space projection remove protected attributes from representations~\cite{ravfogel2020null}; Geiger et al.~\cite{geiger2022causal} use interchange interventions. Concepts often reside in linear subspaces of activations~\cite{rai2024practical,olah2022mechinterp}; attribution and mechanism-level analysis~\cite{parekh2020interpreting,kumar2023essay} motivate auditing which directions drive behavior. ACES applies a subspace lens (extract, stress-test, project-out) to ASR representations and uses project-out as a causal test: if removing the accent subspace reduced disparity, the subspace would be predictively coupled to degradation; we report that it did not.

\section{Method}
\label{sec:method}

\begin{figure}[t]
\centering
\includegraphics[width=\linewidth]{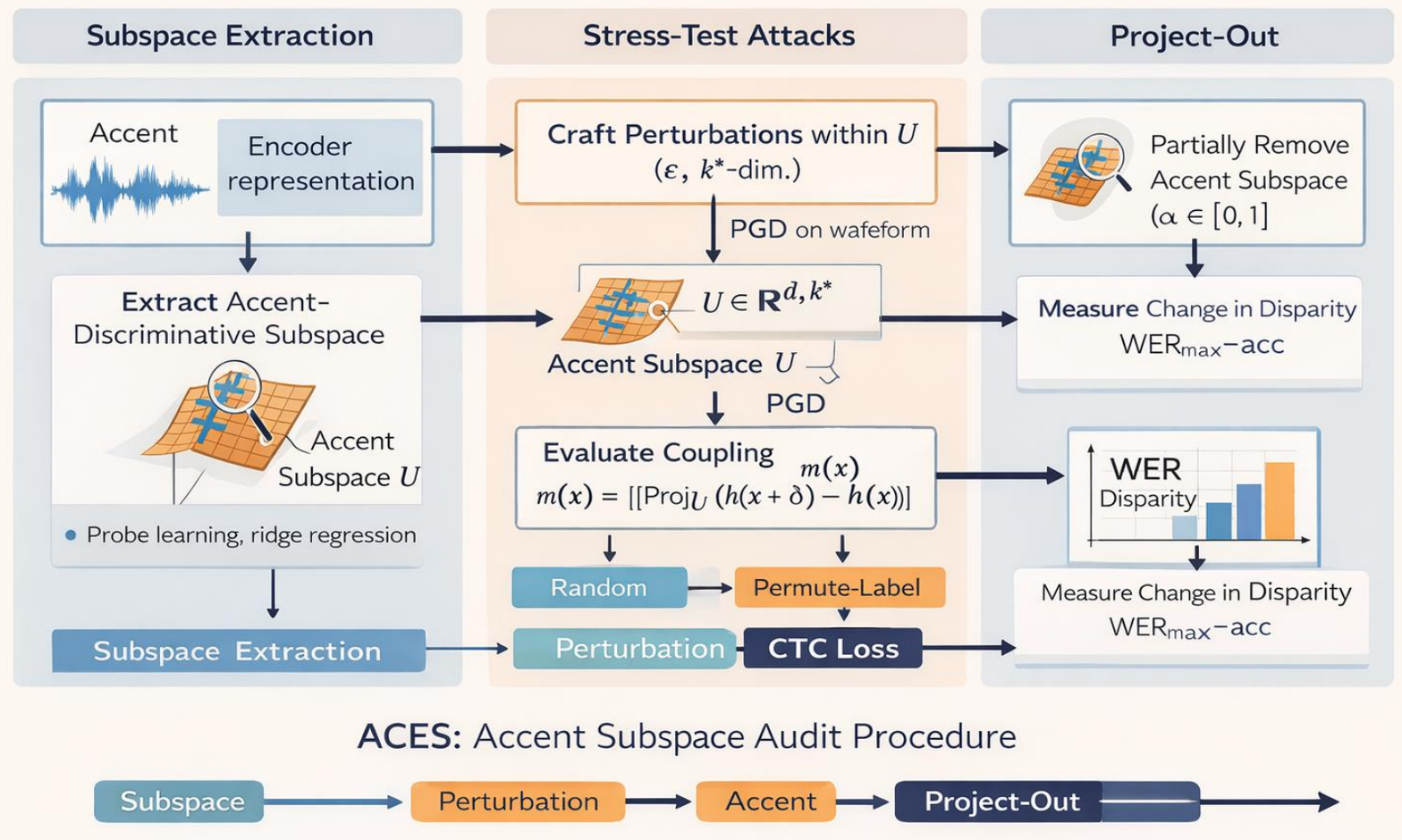}
\caption{ACES: subspace extraction, stress-test (waveform PGD, L2 $\varepsilon$), and project-out. Conceptually, the representation $\mathbf{h}$ is projected onto the plane spanned by $\mathbf{U}$; coupling $m(x)$ measures how much the attack moves $\mathbf{h}$ along that subspace.}
\label{fig:aces_overview}
\end{figure}

We use a single pretrained ASR model (Wav2Vec2-base-960h) and a multi-accent manifest (audio path, accent, text). ACES first extracts an accent subspace and chooses a layer, then stress-tests with subspace-constrained attacks, and finally evaluates project-out at inference. Figure~\ref{fig:aces_overview} summarizes the design and the coupling metric.

\subsection{Subspace extraction and validation}

For each layer $\ell \in \mathcal{L}$ we extract encoder hidden states, mean-pool over time to get utterance embeddings $\mathbf{e}_\ell \in \mathbb{R}^d$, and learn a matrix $\mathbf{U} \in \mathbb{R}^{d \times k}$ whose column span captures accent-discriminative directions. We evaluate several subspace-learning methods (linear probe, LDA, centroid-difference, ridge probe) and report results for the ridge probe, which gave the best validation accuracy; results were qualitatively consistent across methods. We validate each candidate subspace by: (i)~probe accuracy on the projected embeddings versus layer, (ii)~principal angles between subspaces learned on random data halves (stability), and (iii)~correlation between projection onto $\mathbf{U}$ and per-utterance WER. We treat the subspace as an equivalence class and use the projector $\mathbf{P} = \mathbf{U}\mathbf{U}^\top$ for intervention~\cite{ravfogel2020null}. We pick the layer $\ell^*$ that maximizes probe accuracy while keeping stability within a reasonable range (principal angle $\sim$50°).

\subsection{Subspace-constrained attacks}

We perturb the \textbf{waveform} $x$ (normalized to $[-1,1]$): $\delta$ is added to the raw audio, $\varepsilon$ is the L2 norm of $\delta$ in waveform space (we use $\varepsilon = 0.01$, corresponding to ${\sim}60$~dB SNR for a typical 5-second utterance---well below audibility), and we clip $x + \delta$ to $[-1, 1]$, with no preprocessing beyond the model's feature extractor. We use \textbf{untargeted} PGD~\cite{madry2018towards}: we maximize CTC loss~\cite{graves2006ctc} to induce transcription degradation. For accent-subspace and random-subspace attacks we maximize the following objective over $\delta$:
\begin{equation}
\mathcal{L}(\delta) = \mathcal{L}_{\mathrm{CTC}}(x+\delta) + \beta \Vert \Pi_U(\mathbf{h}(x+\delta) - \mathbf{h}(x)) \Vert^2
\label{eq:attack}
\end{equation}
where $\Pi_U = \mathbf{U}\mathbf{U}^\top$ is the projector onto the accent subspace and $\beta \geq 0$ governs the strength of the subspace constraint (larger $\beta$ encourages representation shifts aligned with $\mathbf{U}$). The first term induces transcription degradation; the second encourages $\delta$ that moves $\mathbf{h}$ along the accent subspace. The \textbf{subspace constraint} is thus in the \emph{objective}; the only direct constraint on $\delta$ is the L2 ball of radius $\varepsilon$.

At the same $\varepsilon$ we compare four conditions: clean audio, unconstrained PGD, random-subspace (orthonormal $\mathbf{U}_{\mathrm{rand}} \in \mathbb{R}^{d \times k}$ at $\ell^*$, fixed seed), and accent-subspace ($\mathbf{U}$ from extraction). We define the \textbf{coupling} at layer $\ell^*$ as
\begin{equation}
m(x) = \Vert \Pi_U(\mathbf{h}(x+\delta) - \mathbf{h}(x)) \Vert
\end{equation}
(representation shift along the subspace). A higher mean $m(x)$ for accent-subspace than for random-subspace attacks confirms the perturbation preferentially activates accent-relevant directions. As a specificity control, we repeat the attack comparison using a subspace $\mathbf{U}_{\mathrm{perm}}$ learned on \emph{shuffled} accent labels; if the effect is specific to accent structure, $\mathbf{U}_{\mathrm{perm}}$ should confer no advantage over a random subspace.

\subsection{Project-out intervention}

If accent were largely orthogonal to recognition cues, projecting it out would be expected to reduce disparity. We test this by partially projecting out the accent subspace at $\ell^*$: $\mathbf{e}' = \mathbf{e} - \alpha \mathbf{U}\mathbf{U}^\top \mathbf{e}$, $\alpha \in (0,1]$ (we use $\alpha=0.5$ to attenuate the subspace). We use a forward hook and measure WER and disparity (max $-$ min WER across accents) with and without the hook on clean and on accent-subspace-attacked audio.

\section{Experiments}
\label{sec:experiments}

\subsection{Setup}

\textbf{Model.} We use Wav2Vec2-base-960h~\cite{baevski2020wav2vec2} (12-layer, hidden size 768), fine-tuned on LibriSpeech~\cite{panayotov2015librispeech}. Hidden states are mean-pooled over time for subspace learning, and attacks and hooks use the same encoder at $\ell^*$.

\textbf{Data.} We use the English portion of the Common Voice corpus~\cite{ardila2020commonvoice}, a crowdsourced, open multilingual speech dataset with accent metadata. Our evaluation set is built by \textbf{randomly sampling} utterances from seven accent groups (African, Bermuda, Indian, Malaysia, Singapore, US, Wales) to form a balanced multi-accent subset in the style of AESRC~\cite{accents2020aesrc}: we draw 100 utterances per accent (700 total) from the available validated English recordings, producing a manifest with columns \texttt{audio\_path}, \texttt{accent}, and \texttt{text}. We use speaker-disjoint train/validation splits where speaker IDs are available (e.g., from filenames). Accents may correlate with recording conditions in the source corpus; we do not control for that.

\textbf{Hyperparameters.} For the subspace we sweep layers 2--6 and $k \in \{4,8,16\}$ using multiple probe types (linear probe, LDA, centroid-difference, ridge probe); we report results for the ridge probe, which yielded the best validation probe accuracy, with layer 3 and $k{=}8$ selected by probe accuracy and stability (principal angle $\sim$50°). For attacks we set $\varepsilon = 0.01$ (L2 on waveform), 80 PGD steps, and $\beta = 2.0$ (subspace-constraint strength in Eq.~\ref{eq:attack}); the same $\varepsilon$ and steps are used for unconstrained ($\beta{=}0$), random-subspace, and accent-subspace conditions. For the project-out intervention we use $\alpha = 0.5$. All runs use a fixed random seed for reproducibility.

\textbf{Evaluation.} We report WER and disparity (max $-$ min mean WER across accents, in percentage points)---a range metric aligned with multi-accent benchmarks such as AESRC~\cite{accents2020aesrc} and fairness-focused evaluations~\cite{rai25asrfairbench}. We also report the mean coupling $m(x)$ for each attack condition as a mechanistic diagnostic: a higher mean $m$ for accent-subspace attacks confirms the perturbation preferentially activates accent-relevant directions. Table~\ref{tab:setup} summarizes the full configuration.

\begin{table}[t]
\caption{Setup \& subspace: model, data, $\varepsilon$/steps, chosen $\ell^*$/$\mathbf{U}$, probe acc., stability.}
\label{tab:setup}
\centering
\small
\begin{tabular}{@{}l@{\quad}p{0.48\linewidth}@{}}
\toprule
\textbf{Item} & \textbf{Value} \\
\midrule
Model & Wav2Vec2-base-960h~\cite{baevski2020wav2vec2} \\
Data & Common Voice (en.)~\cite{ardila2020commonvoice}, 7 acc., 100 utt./acc., 700 total \\
$\varepsilon$, steps & 0.01, 80 (waveform L2) \\
Layer $\ell^*$, $k$ & 3, 8 (ridge probe) \\
Probe acc.\ (projected) & 96.3\% \\
Stability (principal angle) & $\sim$50° \\
\bottomrule
\end{tabular}
\end{table}

\subsection{Results}

We report three findings: accent geometry in early layers, gap amplification under subspace-constrained attacks, and the failure of linear project-out as a fairness intervention.

\subsubsection{Accent geometry in early layers}

\begin{figure}[t]
\centering
\includegraphics[width=\linewidth]{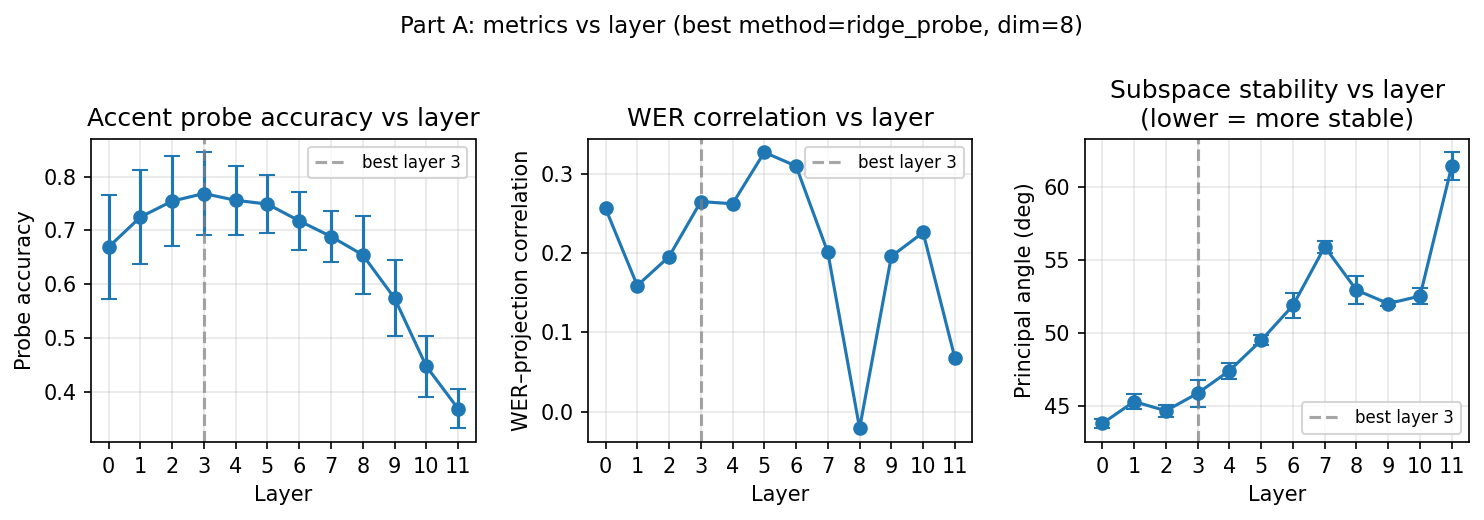}
\caption{Three-track diagnostic (1-column): probe accuracy, corr(projection, WER), stability (principal angle, °) vs.\ layer. Layer~3 maximizes probe accuracy while maintaining stability below 50°; $k{=}8$ (dashed).}
\label{fig:parta_metrics}
\end{figure}

Figure~\ref{fig:parta_metrics} shows the three-track diagnostic (probe accuracy, WER--projection correlation, stability vs.\ layer; stability in principal angle, degrees). Layer~3 maximizes probe accuracy while maintaining subspace stability below 50° principal angle, indicating consistent geometry across splits; we use $k{=}8$. Accent decodability peaks in early layers (2--4) and declines in deeper layers, consistent with prior findings that early wav2vec2 layers encode acoustic features while deeper layers capture linguistic abstractions~\cite{pasad2021layer}; this clarifies why early-layer entanglement with recognition cues matters.

\subsubsection{Gap amplification under subspace-constrained attacks}

Table~\ref{tab:wer} gives mean WER by accent and condition. Disparity rises from 21.3~pp (clean) to 29.7~pp (unconstrained PGD). Unconstrained PGD produces the highest mean WER but \emph{lower} disparity than the subspace conditions because it degrades all accents indiscriminately; the subspace term in Eq.~\ref{eq:attack} redirects perturbation energy toward accent-sensitive directions, yielding more targeted degradation---e.g., US WER rises only to 24.5 (vs.\ 28.4 unconstrained), while Indian WER remains nearly as high (56.3 vs.\ 56.7). Crucially, accent-subspace attacks amplify disparity further (31.8~pp) than random-subspace controls (30.9~pp). Accent-subspace WER exceeds random-subspace WER on all but one accent: the gap is largest for Indian (+2.3~pp), Bermuda (+2.2~pp), and Malaysia (+1.5~pp), moderate for Singapore (+1.3~pp) and US (+1.3~pp), and small for Wales (+0.7~pp). Only African shows a negligible reversal ($-$0.1~pp). While the 0.9~pp gap in overall disparity between accent-subspace and random-subspace is modest, its consistency across accents (6/7 positive, mean per-accent gap ${+}1.3$~pp) and its elimination under the permuted-label control (below) provide converging evidence that the learned subspace captures genuine mechanistic vulnerability structure, not a random artefact. Accents with higher baseline WER suffer larger absolute degradation (Indian: 40.7$\to$56.3; Malaysia: 39.8$\to$52.6), while the lowest-baseline accent (US: 19.4$\to$24.5) is least affected---the attack \emph{widens pre-existing disparities}.

\begin{figure}[t]
\centering
\includegraphics[width=0.85\linewidth]{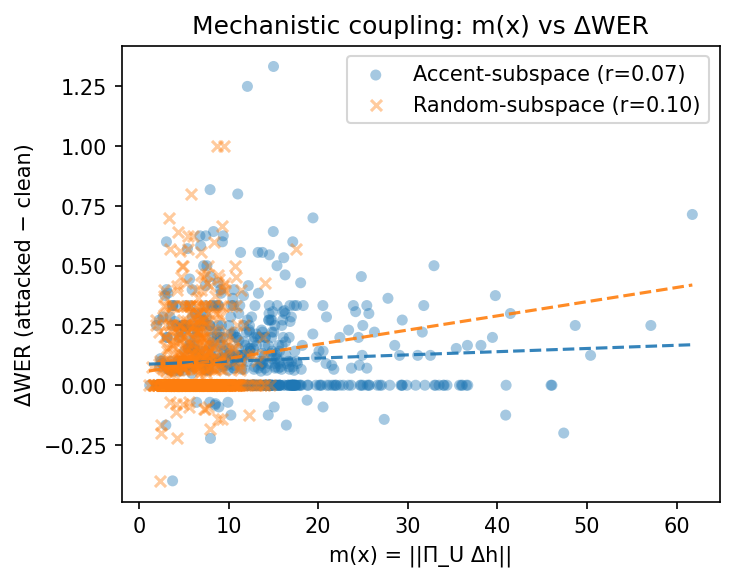}
\caption{Coupling $m(x)$ vs.\ $\Delta\text{WER}$ at layer $\ell^*$. Accent-subspace attacks (blue) produce ${\sim}2\times$ higher mean $m(x)$ than random-subspace (orange), confirming preferential activation of accent directions. Per-utterance $r$ is weak for both due to CTC decoding noise.}
\label{fig:mech_coupling}
\end{figure}

\textbf{Mechanistic coupling.} The mean coupling $m(x) = \Vert \Pi_U \Delta \mathbf{h} \Vert$ is ${\sim}2\times$ higher for accent-subspace attacks (12.2) than random-subspace attacks (6.2), confirming that the accent-subspace perturbation preferentially activates accent-relevant directions in the representation (Figure~\ref{fig:mech_coupling}). Per-utterance correlation between $m(x)$ and $\Delta$WER is weak ($r \approx 0.07$) for both conditions, which we attribute to the discrete, noisy nature of CTC-decoded WER; the aggregate-level separation in $m(x)$ is the robust mechanistic signal.

\textbf{Specificity control (permuted labels).} To rule out the possibility that \emph{any} data-derived subspace would outperform a random one, we learn a subspace $\mathbf{U}_{\mathrm{perm}}$ from the same data but with \emph{shuffled} accent labels, then repeat the attack comparison. With the real accent subspace, accent-subspace WER exceeds random-subspace WER on 6/7 accents (mean gap ${+}1.3$~pp). With $\mathbf{U}_{\mathrm{perm}}$, the gap is centered at zero: only 3/7 accents are positive and the mean gap is ${-}0.5$~pp. This confirms that the advantage is specific to genuine accent structure, not an artefact of using a data-derived subspace.

\begin{figure}[t]
\centering
\includegraphics[width=\linewidth]{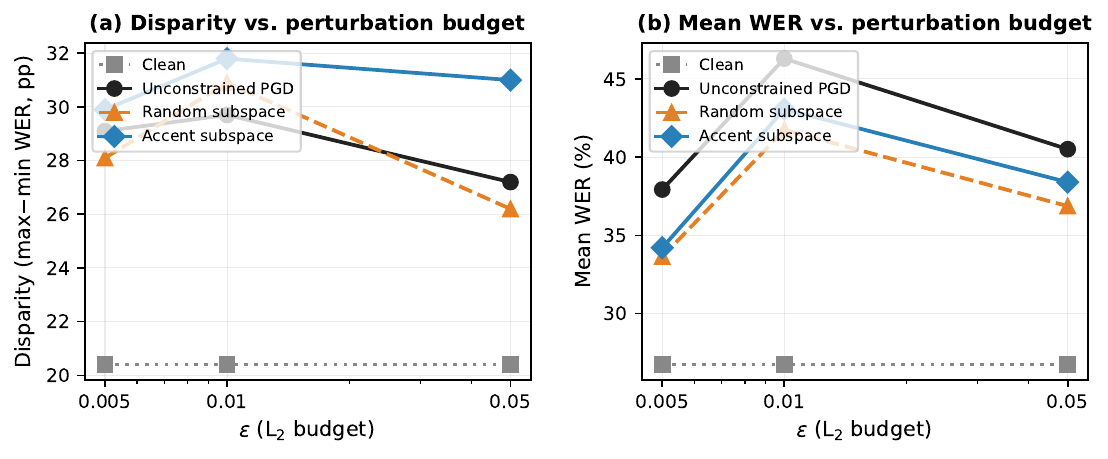}
\caption{Epsilon ablation: disparity (left) and mean WER (right) vs.\ perturbation budget $\varepsilon$. The accent-subspace (blue) maintains high disparity as $\varepsilon$ grows, while the random control (orange) drops off---the gap widens from +1.8~pp ($\varepsilon{=}0.005$) to +4.8~pp ($\varepsilon{=}0.05$).}
\label{fig:eps_ablation}
\end{figure}

\textbf{Epsilon ablation.} Figure~\ref{fig:eps_ablation} sweeps the perturbation budget $\varepsilon \in \{0.005, 0.01, 0.05\}$. At the smallest budget, accent-subspace disparity already exceeds the random control (+1.8~pp). As $\varepsilon$ grows, unconstrained PGD overwhelms all accents equally---random-subspace disparity \emph{drops} from 28.1 to 26.2~pp---while accent-subspace disparity remains elevated (29.9$\to$31.0~pp), widening the gap to +4.8~pp at $\varepsilon{=}0.05$. This confirms that the accent subspace captures genuinely accent-sensitive directions: at larger budgets, random perturbations wash out group differences, but subspace-constrained perturbations maintain targeted disparity amplification.

\begin{table}[t]
\caption{Mean WER (\%) by accent and condition; $\varepsilon{=}0.01$, 80 steps, $\beta{=}2$. Disparity = max $-$ min. Lowest WER per column in \textbf{bold}.}
\label{tab:wer}
\centering
\footnotesize
\begin{tabular}{lcccc}
\toprule
Accent & Clean & Unconstr. & Random & Accent-sub. \\
\midrule
African   & 27.1 & 38.0 & 34.3 & 34.2 \\
Bermuda   & 30.9 & 46.4 & 42.6 & 44.9 \\
Indian    & 40.7 & 56.7 & 54.1 & 56.3 \\
Malaysia  & 39.8 & 58.1 & 51.1 & 52.6 \\
Singapore & 31.9 & 43.6 & 40.8 & 42.2 \\
US        & \textbf{19.4} & \textbf{28.4} & \textbf{23.2} & \textbf{24.5} \\
Wales     & 39.4 & 52.9 & 46.1 & 46.8 \\
\midrule
\rowcolor{gray!20}
Disparity & 21.3 & 29.7 & 30.9 & 31.8 \\
\bottomrule
\end{tabular}
\end{table}

\subsubsection{The failure of linear intervention}

Table~\ref{tab:intervention} shows the effect of partially projecting out the accent subspace ($\alpha{=}0.5$) on both clean and accent-subspace-attacked audio. On clean audio, mean WER rises from 32.7\% to 33.0\% (+0.3~pp) and disparity increases from 21.3 to 21.7~pp. Under attack, mean WER rises from 39.0\% to 39.2\% (+0.3~pp); attacked disparity decreases marginally from 28.4 to 27.9~pp, well within noise. Critically, the degradation is \emph{asymmetric}: high-WER accents suffer the largest clean-WER increases (Wales +0.7~pp, Malaysia +0.6~pp) while the lowest-WER accent (US) is unchanged ($-$0.1~pp). This confirms that accent-discriminative directions overlap with recognition-critical acoustic features; removing them blurs phonetic distinctions and disproportionately affects already fragile accents. Linear project-out is therefore not recommended as a fairness intervention.

\begin{table}[t]
\caption{Project-out intervention ($\alpha{=}0.5$): WER (\%) before (Base) and after (Int.) attenuating the accent subspace on clean and accent-subspace-attacked audio. Attacked WER here uses per-utterance PGD; batch-level aggregation in Table~\ref{tab:wer} yields slightly different values.}
\label{tab:intervention}
\centering
\footnotesize
\begin{tabular}{lcccc}
\toprule
& \multicolumn{2}{c}{Clean} & \multicolumn{2}{c}{Attacked} \\
\cmidrule(lr){2-3} \cmidrule(lr){4-5}
Accent & Base & Int. & Base & Int. \\
\midrule
African   & 27.1 & 27.5 & 31.6 & 31.5 \\
Bermuda   & 30.9 & 30.8 & 41.2 & 42.0 \\
Indian    & 40.7 & 41.0 & 49.0 & 49.0 \\
Malaysia  & 39.8 & 40.4 & 48.4 & 48.4 \\
Singapore & 31.9 & 32.2 & 38.4 & 38.9 \\
US        & \textbf{19.4} & \textbf{19.3} & \textbf{20.6} & \textbf{21.1} \\
Wales     & 39.4 & 40.1 & 43.6 & 43.8 \\
\midrule
\rowcolor{gray!20}
Mean      & 32.7 & 33.0 & 39.0 & 39.2 \\
\rowcolor{gray!20}
Disparity & 21.3 & 21.7 & 28.4 & 27.9 \\
\bottomrule
\end{tabular}
\end{table}

\section{Discussion and Limitations}
\label{sec:discussion}

ACES assumes that accent is encoded in an approximately linear subspace; if the structure were strongly nonlinear, other tools would be needed. We interpret the project-out result via an \textbf{acoustic--phonetic overlap} hypothesis: the $k{=}8$ subspace likely contains directions that both differentiate accents and distinguish phonemes; removing them muddies phonetic boundaries and hits accents already near the failure threshold harder. The asymmetry in Table~\ref{tab:intervention}---Wales (+0.7~pp) and Malaysia (+0.6~pp) vs.\ US ($-$0.1~pp) on clean audio---is consistent with a threshold effect: accents closer to the model's failure boundary lose more from the loss of phonetic resolution. The consistency of accent-subspace advantage across nearly all tested accents (6/7) in the attack experiment, including both native (Wales) and non-native (Indian, Malaysian, Singaporean) English varieties, together with the permuted-label control (which eliminates the advantage), strengthens the claim that the learned subspace captures real mechanistic vulnerability structure specific to accent encoding. In practice, ACES can be used to: (i)~detect whether accent directions predict degradation; (ii)~evaluate whether mitigation strategies reduce coupling; (iii)~flag unsafe erasure interventions. We recommend running ACES to audit models before deploying in fairness-sensitive applications. A deeper qualitative understanding of \emph{what} is erased is left for future work. We use max$-$min disparity for interpretability; variance-based measures yield qualitatively similar patterns but are less intuitive for small accent counts. Project-out may also create out-of-distribution representations for later layers. We evaluate one model and seven accents; single-model evaluation is standard in mechanistic interpretability studies, and the ACES methodology is model-agnostic---extending to larger architectures (e.g., Whisper~\cite{radford2023whisper}, HuBERT~\cite{hsu2021hubert}) would strengthen generalizability.

\section{Conclusion}

ACES uses accent subspaces as audit instruments: subspace extraction, subspace-constrained stress tests (waveform PGD, matched L2), and project-out intervention. On Wav2Vec2-base with seven English accents, accent-subspace-constrained attacks amplify the WER disparity gap from 21.3 to 31.8~pp, exceeding random-subspace controls (30.9~pp) on the majority of accents (6/7); a permuted-label control confirms this advantage is specific to accent structure. Accents with higher baseline WER suffer larger absolute degradation, confirming that the learned subspace captures directions along which the model is most fragile. Project-out ($\alpha{=}0.5$) worsens clean WER (+0.3~pp mean) and clean disparity (+0.4~pp) while leaving attacked disparity essentially unchanged ($-$0.5~pp), with the degradation concentrated on high-WER accents; subspaces are useful for auditing rather than as a standalone fairness fix.

\section{Generative AI Use Disclosure}

Generative AI tools were used for programming, editing, and polishing the manuscript.

\bibliographystyle{IEEEtran}
\bibliography{mybib}
\end{document}